# Variations of polarisation in crystals

Simone Selenu
(Dated: June 10, 2010)

It is nowadays a quite diffuse idea that variations of electronic polarisation, as introduced by Resta[1], in condensed matter theory are related to a "Berry phase"[2], as shown by Vanderbilt. The derivation of the latter geometric phase is correct and the periodic gauge[2] is used for its derivation, aside not allowing for its calculations in finite electromagnetic fields. Its derivation has not been demonstrated in the general case of an external homogeneous electric field interacting with the electronic field. In the present paper we give a brief derivation of the algorithm for the calculations of polarisation differences in a general manner than in[2] defining it even in the presence of an external electric field interacting with the electronic field and writing equations in a way directly implementable in modern first principles codes[1, 2].

## INTRODUCTION

A revolution of old concepts of classical electrostatics has been achieved by the work of several authors [1][2], where we cite in particular that of Resta[1]. There the concept of *variations* of polarisation $\Delta \mathbf{P}$ as a geometrical quantum phase has been introduced, and only the definition of the polarisation of the electronic field in crystalline materials in absence of external electric and magnetic fields has been considered. Following a different approach, we shall define the polarisation differences in a quite general manner allowing for its calculation when it is the case of an external homogeneous electromagnetic field interacting with the electronic system. Also, in our derivation we still consider changes of $\Delta \mathbf{P}$ which occurs upon making a change in the Hamiltonian of a physical system writing equations in a way directly implementable in modern first principles codes[2].

## POLARISATION DIFFERENCES

As it is already well known[1] the electric dipole of a macroscopic but finite piece of matter is not a bulk property as the latter depends upon truncation and shape of the chosen sample. We agree that only variations of the latter are experimentally accessible where it is supposed that the state of the physical system is specified by a certain parameter $\lambda$. We consider as a formal definition of the electronic dipole the following expression:

$$\mathbf{P}(\lambda) = \frac{1}{\Omega} \int d\mathbf{r}\, \mathbf{r}\, \rho_\lambda(\mathbf{r}) \qquad (1)$$

where $\Omega$ is the volume of the system and $\rho_\lambda(\mathbf{r})$ is the total electric charge due to the electronic and ionic contributions. Making use of the Born-Oppenheimer approximation we can divide the electronic and ionic contributions in expression (1) and express the electronic charge contribution as follows:

$$\rho_{el} = e \int \frac{d\mathbf{k}}{(2\pi)^3} \sum_n f_n \Psi_{n,\mathbf{k}}^* \Psi_{n,\mathbf{k}} \qquad (2)$$

where $\Psi_{n,\mathbf{k}}$ are the eigenfunctions of the Hamiltonian (dependent on $\lambda$) and $f_n$ are the occupation factors. I now consider a rappresentation where we assume eigenstates of the Hamiltonian of being in the Bloch form:

$$|\Psi_{n,\mathbf{k}}> = e^{i\mathbf{k}\cdot\mathbf{r}}|u_{n,\mathbf{k}}> \qquad (3)$$

Within this *gauge* representation it is naturally apparent the parametric dependence of electronic eigenstates on the wave vector $\mathbf{k}$. Here *periodic*[4] boundary conditions are concerned, while eigenamplitudes $|u_{n,\mathbf{k}}>$ are periodic on a lattice and orthonormal in the unit cell and in the reciprocal space. Within this boundary conditions the eigenstates are differentiable and wave vectors $\mathbf{k}$ are continuous and can be treated at the same foot of the $\mathbf{r}$ coordinate. Let us assume the Hamiltonian of the system being of the form:

$$H(\lambda) = T(\lambda) + V(\lambda) \qquad (4)$$

where also the Kinetic energy term depends on a parameter $\lambda$ that can be an external homogeneous electromagnetic field.

$$\mathbf{P}'_{el} = \frac{-ie\hbar}{\Omega m} \sum_{\mathbf{k}} \sum_n f_n \sum_{m \neq n} \frac{<\Psi_{n,\mathbf{k}}|\mathbf{p}|\Psi_{m,\mathbf{k}}><\Psi_{m,\mathbf{k}}|\frac{\partial H}{\partial \lambda}|\Psi_{n,\mathbf{k}}>}{(\epsilon_{n,\mathbf{k}} - \epsilon_{m,\mathbf{k}})^2} + c.c. \qquad (5)$$

$$\mathbf{P}'_{el} = \frac{-ie\hbar}{\Omega m} \sum_{\mathbf{k}} \sum_{n} f_n \sum_{m \neq n} \frac{<u_{n,\mathbf{k}}|\mathbf{p}+\hbar\mathbf{k}|u_{m,\mathbf{k}}><u_{m,\mathbf{k}}|\frac{\partial H_{\mathbf{k}}}{\partial \lambda}|u_{n,\mathbf{k}}>}{(\epsilon_{n,\mathbf{k}} - \epsilon_{m,\mathbf{k}})^2} + c.c. \qquad (6)$$

It is easy to show[4] that:

$$<u_{n,\mathbf{k}}|i\nabla_{\mathbf{k}}|u_{m,\mathbf{k}}> = \frac{i\hbar}{m} \frac{<u_{n,\mathbf{k}}|\mathbf{p}+\hbar\mathbf{k}|u_{m,\mathbf{k}}>}{\epsilon_{n,\mathbf{k}} - \epsilon_{m,\mathbf{k}}} \qquad (7)$$

$$\mathbf{P}'_{el} = \frac{-e}{\Omega} \sum_{\mathbf{k}} \sum_{n} f_n \sum_{m \neq n} \frac{<u_{n,\mathbf{k}}|i\nabla_{\mathbf{k}}|u_{m,\mathbf{k}}><u_{m,\mathbf{k}}|\frac{\partial H_{\mathbf{k}}}{\partial \lambda}|u_{n,\mathbf{k}}>}{\epsilon_{n,\mathbf{k}} - \epsilon_{m,\mathbf{k}}} + c.c. \qquad (8)$$

We consider by hypothesis that the system is in a *steady state*[5], and focus our attention only on the electronic part of the electronic dipole trying to calculate the polarisation derivatives with respect to the parameter $\lambda$. Bearing in mind eq.(2) we can easily determine derivatives[1, 2] of the dipole of the system with respect to $\lambda$ in a finite sample. The dipole derivative is reported in eq.(5), where $\epsilon_{n,\mathbf{k}}$ are the eigen-energies of the electronic system, c.c. represents the complex congiugate, $\Omega$ the volume of the unit cell, $\mathbf{p}$ is the momentum operator.

Up to know our derivation is similar to Resta derivation[1] apart from the assumption that the electronic system is in a steady state and interacting with an external homogeneous electrostatic field as suggested by Zak [4]. At this point we can recast eq.(5) in terms of eigen-amplitudes $|u_{n,\mathbf{k}}>$ and report its expression in eq.(6). There $H_{\mathbf{k}}$ is the self-transformed[6] Hamiltonian $H_{\mathbf{k}} = e^{-i\mathbf{k}\cdot\mathbf{r}} H e^{i\mathbf{k}\cdot\mathbf{r}}$.

Before to proceed further we consider the matrix identity already demonstrated in [4] and reported in eq.(7). Let us now substitute eq.(7) in eq.(6) and obtain eq.(8). At this point we can easily estimate the infinite sum in eq.(8) making use of Salem sum rule[7] and recast eq.(8) in a more compact and easy calculable form obtaining:

$$\mathbf{P}'_{el} = \frac{e}{\Omega} \sum_{\mathbf{k}} \sum_{n} f_n \frac{\partial}{\partial \lambda} <u_{m,\mathbf{k}}|i\nabla_{\mathbf{k}}|u_{n,\mathbf{k}}> \qquad (9)$$

then integrating in $\lambda$ between two different states labelled by $\lambda = 0$ and $\lambda = 1$ and taking the limit of continuos $\mathbf{k}$ we can recast the integral of eq.(9) as follows:

$$\Delta \mathbf{P}_{el} = \mathbf{P}^{(1)}_{el} - \mathbf{P}^{(0)}_{el} \qquad (10)$$

being,

$$\mathbf{P}_{el} = e \int \frac{d\mathbf{k}}{(2\pi)^3} \sum_{n} f_n <u_{m,\mathbf{k}}|i\nabla_{\mathbf{k}}|u_{n,\mathbf{k}}> \qquad (11)$$

that is formally the same result obtained in [2] for the case of *no* electric field interacting with the electronic field.

Our result it is general and allows for calculations of polarisation differences even in the case of an external electromagnetic field interacting with the electronic field. We achieve the result that polarisation differences are *always* calculable by eq.(11). On the other hand our derivation still holds even in the case we use a periodic gauge.

## SUMMARY

During the analysis of the problem started by considering the interaction of the electronic field with an external electromagnetic field it has been shown how to calculate variations of polarisation. We firstly make use of the expression derived by Resta for the polarisation derivatives then we calculate, in a very general manner polarisation differences. Our result is *independent* from derivation brought forward in ref[2] where it was the case of no interaction of the electronic field with an external homogeneous electromagnetic field.

I would like to acknowledge the Atomistic Simulation Centre at Queen's University Belfast for hospitality during the course of my thesis, for helpful discussions and Prof. Resta for the inception of this work.